\documentclass[useAMS,usenatbib,letterpaper]{mn2e}
\pdfoutput=1

\usepackage{graphicx}
\usepackage[totalwidth=480pt,totalheight=680pt,layoutvoffset=0.5cm]{geometry}
\usepackage{color}
\usepackage{hyperref}

\newcommand{\apj}{ApJ}
\newcommand{\apjs}{ApJ Suppl.}
\newcommand{\mnras}{MNRAS}

\newcommand{\aap}{A{\&}A}

\newcommand{\apjl}{ApJL}

\newcommand{\changed}[1]{#1} 
\newcommand{\eq}[1]{(\ref{eq:#1})} 
\newcommand{\eqq}[1]{Eq.~(\ref{eq:#1})}

\begin{document}

\title[Photo-$z$]{Improving photometric redshifts with Ly$\alpha$ tomography}

\author[Schmittfull \& White]{Marcel Schmittfull$^{1}$,
Martin White$^{1, 2, 3}$\\
$^1$ Berkeley Center for Cosmological Physics, University of California,
Berkeley, CA 94720, USA \\
$^2$ Department of Astronomy and Department of Physics,
University of California, Berkeley, CA 94720, USA \\
$^3$ Lawrence Berkeley National Laboratory, 1 Cyclotron Road, Berkeley, CA
93720, USA
}
\date{\today}
\pagerange{\pageref{firstpage}--\pageref{lastpage}}

\maketitle

\label{firstpage}

\begin{abstract}
Forming a three dimensional view of the Universe is a long-standing
goal of astronomical observations, and one that becomes increasingly
difficult at high redshift.
In this paper we discuss how tomography of the intergalactic medium
(IGM) at $z\simeq 2.5$ can be used to estimate the redshifts of massive
galaxies in a large volume of the Universe based on spectra of galaxies
in their background.  Our method is based on the fact that hierarchical
structure formation leads to a strong dependence of the halo density on
large-scale environment.  A map of the latter can thus be used to refine
our knowledge of the redshifts of halos and the galaxies and AGN which
they host.  We show that tomographic maps of the IGM at a resolution of
$2.5\,h^{-1}$Mpc can determine the redshifts of more than 90 per cent of
massive galaxies with redshift uncertainty $\Delta z/(1+z)=0.01$.
Higher resolution maps allow such redshift estimation for lower mass
galaxies and halos. 
\end{abstract}


\section{Introduction}
\label{se:intro}

Measuring accurate distance for extragalactic objects is one of the most
challenging problems in observational cosmology.  Distances are needed in
order to properly map large-scale structure, to convert observed into
intrinsic properties and to correctly situate objects within the cosmic web.
The highest quality distance estimates for high redshift objects come from
spectroscopy, which can allow accurate measurements of redshift if suitably
high signal-to-noise spectra are available.
An estimate of the redshift can also be obtained directly from the photometry
\citep[a ``photo-$z$''; e.g.~see][for recent reviews]{Hil10,Dah13,San14,Rau15},
though typically with lower precision and a higher catastrophic error rate.

A particularly relevant example is situating objects within the COSMOS field
\citep{Sco07,Cap07}, which has become a premier extragalactic field for a wide
variety of studies.
Fully exploiting this deep sky map requires information on the redshifts of
the objects, and the wide wavelength coverage and numerous bands available
in COSMOS leads to very good photo-$z$ performance
\citep[near $0.03-0.06$ in $\Delta z/(1+z)$ for bright galaxies at $z=2.5$;][]{Ilb13,Lai16}.
Even so the implied line-of-sight resolution is relatively
poor\footnote{An fractional redshift uncertainty of $\Delta z/(1+z)$
translates into a distance uncertainty of
$\delta\chi=[c(1+z)/H(z)]\ \Delta z/(1+z)$ with $c(1+z)/H(z)\simeq 4\,$Gpc
at $z=2.5$.} ($>100\,h^{-1}$Mpc at $1\,\sigma$ for $z\simeq 2.5$)
and accurate redshifts are easier to obtain for some types of galaxies
than others.

In this paper we discuss how knowledge of the large-scale environment of
galaxies, as traced by fluctuations in neutral hydrogen absorption by the
intergalactic medium, can be used to improve photo-$z$ performance.
In particular we address how Ly$\alpha$ forest tomography
\citep{Cau08,Cai14,Lee14a,Lee14b}
can be used to improve the redshift estimates of massive galaxies, in many
cases significantly.
We shall take as a test case simulated data such as would be returned by the
``COSMOS Lyman-Alpha Mapping And Tomography Observations'' (CLAMATO) survey,
which will cover 1 deg$^2$ within the COSMOS field.
By sampling the IGM absorption along and across sightlines with Mpc
spacing, CLAMATO allows tomographic reconstruction of the 3D Ly$\alpha$ forest
flux field. These tomographic maps have a \changed{line of sight} resolution similar to 
the average
transverse sightline spacing and naturally avoid projection effects or redshift
errors.  The final CLAMATO survey would provide a tomographic map with a
volume of roughly $70\times 70\times 230\,h^{-1}$Mpc at $2.5\,h^{-1}$Mpc
resolution.  In such a map one can easily locate large overdensities
\citep{Sta15a}, voids \citep{Sta15b} and see the cosmic web \changed{including its filaments and sheets} \citep{LeeWhi16}.

\changed{In structure formation from gravitational instability}
most of the volume of the Universe is
underdense, while the halos which host galaxies and AGN preferentially live
in overdense regions, with the tendency \changed{being} the strongest for the most massive
halos.  This simple fact allows us to significantly improve the performance
of photo-$z$s given knowledge of the large-scale density field as traced by
the Ly$\alpha$ forest.
Though our method is different in detail, the end goal and basic insight
is similar to \citet{Kov10}, \changed{\citet{Rak11},} \citet{JasWan12} 
and in particular \citet{Ara15}.

The outline of the paper is as follows.
In \S\ref{sec:sim} we introduce the N-body simulation which we use to test and
illustrate our method, which we motivate and describe in \S\ref{sec:method}.
Our main results are presented in \S\ref{sec:results}, while \S\ref{sec:future}
describes directions for future research which could further improve the
performance of redshift determination using IGM tomography.
Finally we conclude in \S\ref{sec:conclusions}.

\section{Simulations}
\label{sec:sim}

In order to demonstrate the \changed{potential} of our method we make use of a set of
mock catalogs based on N-body simulations.  The simulations are described
in more detail in \citet{Sta15a,Sta15b,Lee16}.
Briefly, the mock catalogs are generated from a high-resolution N-body
simulation which employed $2560^3$ equal mass
($8.6 \times 10^7 \, h^{-1} M_\odot$)
particles in a $256\,h^{-1}$Mpc periodic, cubical box leading to a
mean inter-particle spacing of $100\,h^{-1}$kpc.
The assumed cosmology was of the flat $\Lambda$CDM family, with
$\Omega_{\rm m}\simeq 0.31$, $\Omega_{\rm b} h^2\simeq 0.022$,
$h=0.6777$, $n_s=0.9611$, and $\sigma_8 = 0.83$, in agreement with
\citet{Planck2013}.
We shall work throughout with the $z=2.5$ output of the simulation, for
which we have halo catalogs and mock Ly$\alpha$ forest data.
The Ly$\alpha$ forest was simulated using the fluctuating Gunn-Peterson
approximation which is sufficient to model the the large-scale features
in the IGM at $z \simeq 2-3$
\citep[see e.g.][]{MeiWhi01,Ror13}.
\changed{We model the impact of observational noise by smoothing the simulated flux field, noting that larger noise leads to lower resolution of the reconstructed
flux maps.}
\changed{Throughout our paper we account for redshift space distortions by
moving halos by their line-of-sight velocity and by using the redshift space 
Ly$\alpha$ flux generated from the simulations of \citet{Sta15a,Sta15b,Lee16}.
}

It is not our intention to provide a detailed modeling of galaxy formation
within the simulation, but we anticipate that massive galaxies at $z\simeq 2.5$
should live at the centers of the most massive dark matter halos at the
same era. 
\changed{We choose a stellar mass limit of $3\times 10^{10}\,M_\odot$ based on highly complete samples of galaxies in COSMOS \citep[see e.g.][Fig.~5]{Muz13}.
Using the conversion of \citet{Mos13} to halo mass at $z\simeq 2.5$ 
and taking into account scatter in the stellar-mass--halo-mass relation
suggests taking halos more massive than $10^{12}\,h^{-1}M_\odot$ as a proxy for ``massive'' galaxies in the COSMOS field.}

\section{Method}
\label{sec:method}

Hierarchical structure formation in cold dark matter models leads to a
strong dependence of the halo mass function upon the large-scale density.
In regions where the density is larger than average the number density of
massive halos is increased, while in regions where it is smaller than
average the number density of massive halos is decreased -- often quite
dramatically \citep{ColKai89,MoWhi96,TinCon09}.

It is easy to see this within the Press-Schechter formalism
\citep{PreSch74,BBKS,Pea99}
where the number of halos is related to the number of peaks in the
smoothed initial density field which exceed a threshold, $\delta_c$.
In the presence of a long-wavelength perturbation the small-scale
fluctuations need an amplitude of $\delta_c-\delta_{\rm long}$ in order
to form a halo.  This is more common in overdense regions and less common
in underdense regions, with the amplitude of the effect being larger for
more massive halos.  Indeed, within this peak-background split argument,
the large-scale bias of halos is the fractional change in the number density
of halos per infinitesimal change in $\delta_{\rm long}$.
For the Press-Schechter mass function this bias is $1+(\nu^2-1)/\delta_c$,
where $\nu=\delta_c/\sigma(M)$ is the number of $\sigma$ a fluctuation has
to be in order to cross $\delta_c$.  Since larger halos correspond to a
larger smoothing scale and smaller $\sigma$, the more massive halos are more
biased and their number density is more sensitive to being in an overdense or
underdense region.  The halos of interest to us here are all on the
exponential tail of the mass function ($\nu\gg 1$) and are highly biased
\changed{tracers of the dark matter, typically in clusters.}

Since the Ly$\alpha$ flux tracks the large-scale density we expect that
massive halos will preferentially reside in regions of negative $\delta_F$,
as we see in Table \ref{tab:fractions}.
Since such extrema occupy only a small fraction of the volume (much of
the volume is occupied by voids) we can limit the positions of halos
\citep[see also][for related ideas]{Kov10,JasWan12,Ara15}.

Within the Press-Schechter formalism with the peak-background split,
the number density of rare, highly biased halos is a Gaussian in the
threshold, $\delta_c$.  If the flux overdensity is a linearly biased
tracer of the matter field, we thus expect the number density of halos
to scale as $\exp[-a\delta_F-b\delta_F^2]$. 

\begin{table}
\begin{center}
\begin{tabular}{cccccc}
&& \multicolumn{4}{c}{lg$M_{\rm min}$} \\
$\delta_F^{\rm lim}$ & Vol
    & $ 11$ & $ 12$ & $ 13$ & $ 14$ \\ \hline
 0.00 &    44.45 &  91.9 &  99.5 &   100 &   100 \\
-0.15 &     6.64 &  39.0 &  67.2 &  99.7 &   100 \\
-0.30 &     0.58 &   6.7 &  17.1 &  72.3 &   100 \\
-0.45 &     0.03 &   0.4 &   1.4 &  12.9 &   100
\end{tabular}
\end{center}
\caption{The fraction (in per cent) of the volume and of the halos
more massive than $M_{\rm min}$ (in $h^{-1}M_\odot$) that lie in
regions of the simulation with $\delta_F$ (smoothed with a Gaussian
of $2.5h^{-1}$Mpc) less than $\delta_F^{\rm lim}$.
We see that more massive halos live preferentially in regions
of lower $\delta_F$, even though those regions occupy a very
small fraction of the total volume.}
\label{tab:fractions}
\end{table}

\begin{figure*}
\begin{center}
\resizebox{0.9\textwidth}{!}{\includegraphics{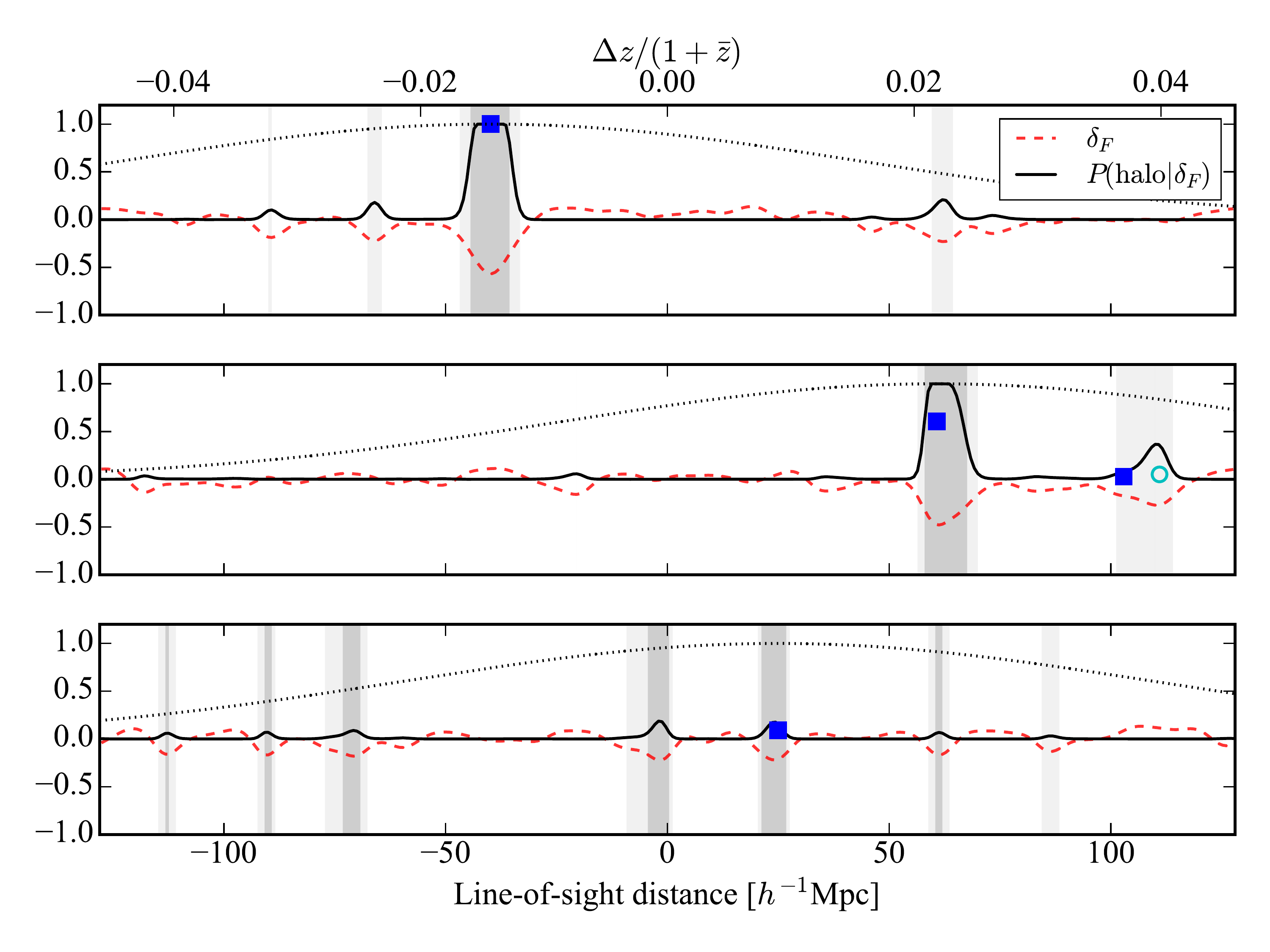}}
\end{center}
\caption{Three sample lines-of-sight through the simulation, showing the
Ly$\alpha$ flux (smoothed with a $2.5\,h^{-1}$Mpc Gaussian; red dashed),
the probability of the halo being at each position (black solid; normalized
to peak at unity in the top panel) and the locations of halos
($M>10^{12}\,h^{-1}M_\odot$) within $r<0.5\,h^{-1}$Mpc (solid blue squares)
or $0.5\,h^{-1}\mathrm{Mpc}<r<1\,h^{-1}\mathrm{Mpc}$ (open cyan circles) 
of the line-of-sight.
The bottom $x$-axis gives the (comoving) distance while the upper panel
shows the equivalent redshift offset from the central redshift $\bar z=2.5$ 
of the skewer.
Shaded regions show credible redshift regions (highest posterior density
intervals) predicted from $p(z)$, assuming 90 per cent (light grey) or 
68 per cent (dark grey) confidence level.
The vertical offset of the squares and circles representing the halos is 
proportional to the logarithm of their mass minus 12.
We have plotted skewers passing through the most massive
(top; $10^{14}\,h^{-1}M_\odot$), $100^{\rm th}$ most massive
(middle; $1.6\times 10^{13}\,h^{-1}M_\odot$) and $10,000^{\rm th}$
(bottom; $1.5\times 10^{12}\,h^{-1}M_\odot$) in the simulation at $z=2.5$.
In the lower panel the significance of the detection is low, and the large
number of peaks is because there are many regions along the skewer where
the flux density is similar to that where the true halo resides.
The dotted lines show a Gaussian of width $0.03$ in $\Delta z/(1+\bar z)$ as an
indication of how well a good photometric redshift would constrain $p(z)$.
}
\label{fig:skewers}
\end{figure*}

We have estimated the conditional probability of seeing a halo
(more massive than $10^{12}\,h^{-1}M_\odot$)
given the smoothed Ly$\alpha$ flux directly from the simulations.
Along any line of sight we have the smoothed Ly$\alpha$ flux field,
$\delta_F$.  To estimate the probability that a given galaxy (halo)
will lie at a given redshift we use Bayes theorem.  Specifically we
know the distribution of $\delta_F$ in the simulation, and we know
the distribution of $\delta_F$ at the halo locations.  Then
\begin{equation}
\label{eq:HistRatio}
  P({\rm halo}|\delta_F) = \frac{P(\delta_F|{\rm halo})}{P(\delta_F)}
\end{equation}
In the simulations we can estimate this as the ratio of two histograms:
\changed{First, the histogram of the flux density in the vicinity of massive halos;
second, the histogram of the flux density over the whole simulated 
volume.}
The result is a monotonically decreasing function of $\delta_F$ which
can be well fit by a Gaussian (as expected from the arguments above)
\begin{equation}
\label{eq:pcondfit}
  P({\rm halo}|\delta_F)\propto e^{-a\delta_F-b\delta_F^2}
\end{equation}
with two parameters ($a$ and $b$) aside from the normalization.  To
obtain this fit we compute the histogram ratio \eq{HistRatio} from our
simulations using all halos more massive than $10^{12}\,h^{-1}M_\odot$
and the flux density interpolated to a $320^3$ grid and smoothed with
a Gaussian kernel $W_R(r)=\exp[-r^2/2R^2]$ with smoothing scale
$R=2.5\,h^{-1}$Mpc.  This gives $a=32.3$ and $b=37.1$ if we restrict
the fit to smoothed flux values $\delta_F>-0.6$ that are present in
the simulation.

Our procedure to predict the redshift PDF for galaxies given the flux
along their lines of sight is thus as follows.  Estimate the smoothed Ly$\alpha$
flux field, as described in \citet{Lee14b,Sta15a}.  Along the line of
sight to each galaxy, compute $p(z)=P\left({\rm halo}|\delta_F(z)\right)$
using \eqq{pcondfit} with the best-fit values for $a$ and $b$ quoted above.
This redshift PDF along the line of sight of each galaxy 
is the final output of our method.

\section{Results}
\label{sec:results}

\begin{figure}
\begin{center}
\resizebox{\columnwidth}{!}{\includegraphics{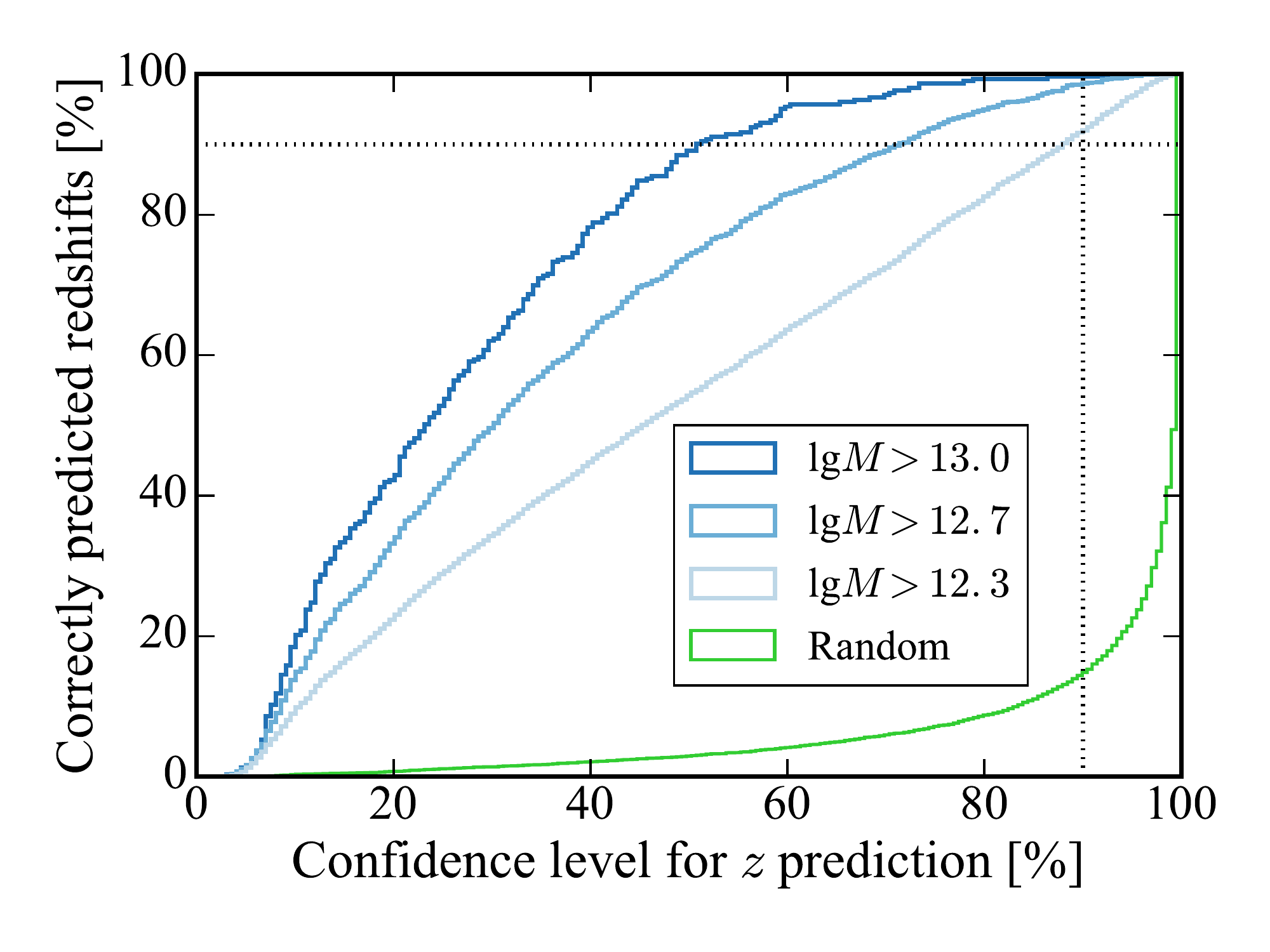}}
\end{center}
\caption{Cumulative fraction of correctly predicted halo redshifts
as a function of the confidence level used to select
credible redshift regions. This quantifies the error rate of predicted
halo redshifts.
The plot is obtained as follows:
For each halo we compute $p(z)$ along the line of sight using \eqq{pcondfit}.
We then ask what is the lowest confidence level $x$ that we would have to
choose so that the credible redshift region based on that 
confidence level includes the true halo redshift. 
If the halo is in a high
confidence region, this value $x$ would be very small.  If the halo is in
a low likelihood region we would need a very conservative confidence
level (high $x$) for the credible redshift region to contain the halo 
and this region would encompass most of the probability.
The plot shows a
cumulative histogram of confidence levels $x$ obtained from all halos
above a certain mass (see legend). The vertical axis
therefore corresponds to the fraction $y$ of halos for which the confidence level
$x$ specified on the horizontal axis is good enough to correctly
predict the redshift of $y$ per cent of the halos.
The random curve is obtained by putting halos in random locations.
}
\label{fig:quantiles}
\end{figure}

\begin{figure}
\begin{center}
\resizebox{\columnwidth}{!}{\includegraphics{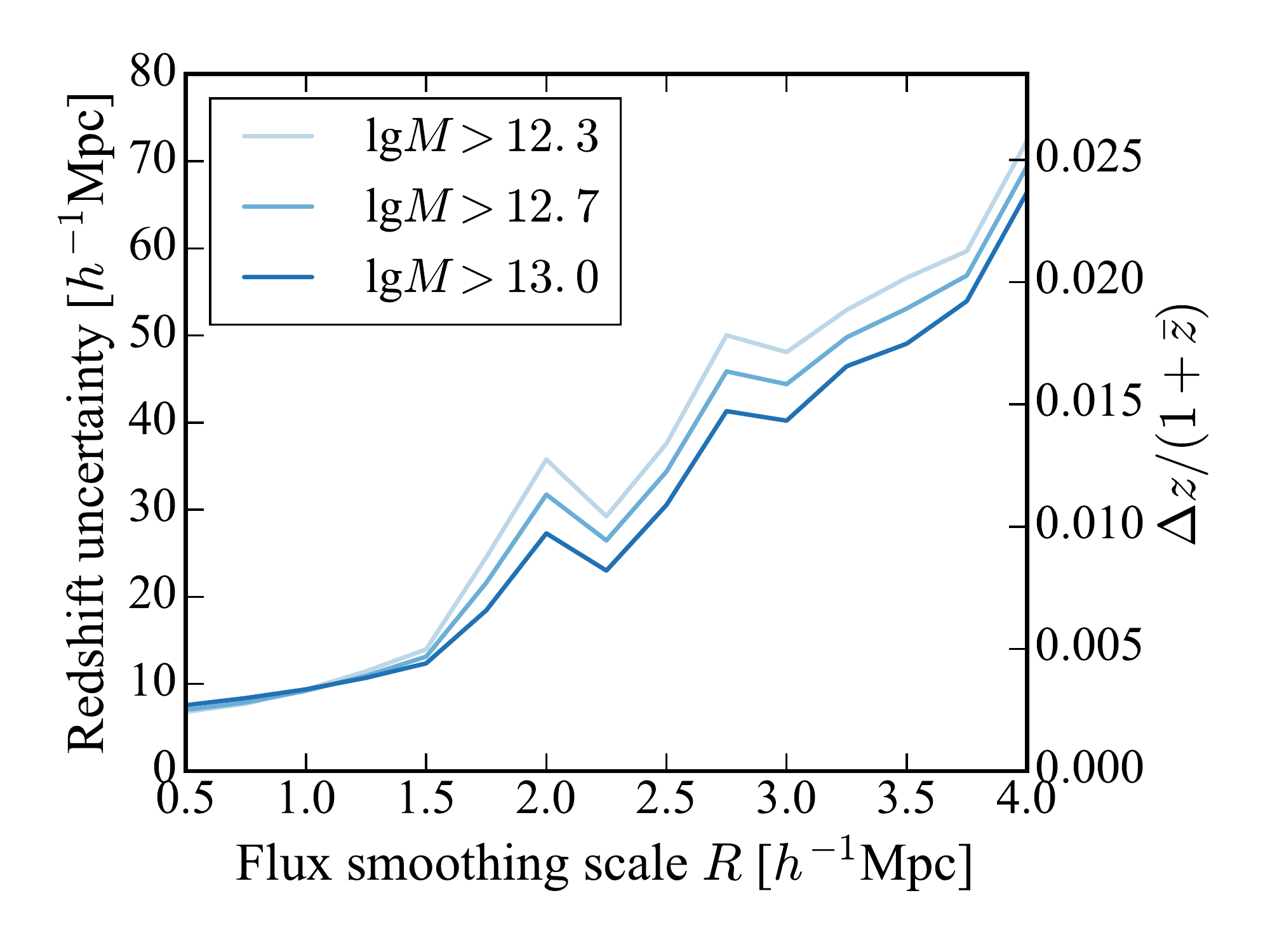}}
\end{center}
\caption{Uncertainty of predicted line-of-sight position or redshift,
as a function of flux smoothing scale $R$.
Specifically, we define the redshift uncertainty as the total width 
of $90$ per cent
credible intervals for predicted redshifts averaged over all skewers
containing at least one halo above the minimum mass specified in the legend.
If the predicted $p(z)$ is multi-modal, we quote the summed width of all
disconnected intervals (not the distance between $p(z)$ peaks).}
\label{fig:credible_R}
\end{figure}

Fig.~\ref{fig:skewers} gives three examples of redshift PDFs obtained from
the simulation.  The three skewers were chosen to be the lines of sight to the
most massive, $100^{\rm th}$ most massive and $10,000^{\rm th}$ most massive
halos in the simulation, with masses ranging from $10^{14}\,h^{-1}M_\odot$
to $10^{12}\,h^{-1}M_\odot$.  The dashed red line shows the flux while the
solid black line shows $p(z)$ (normalized to peak at unity in the top panel).
The squares and circles show the positions of halos within $0.5$ and 
$1\,h^{-1}$Mpc (comoving) of the
line of sight, with vertical position indicating their mass.  
The dotted line shows a Gaussian (normalized to peak at
unity) centered at the true position of the halo with width
$\Delta z/(1+z)=0.03$, comparable to a good photo-$z$.
Clearly including prior photo-$z$ information could serve to eliminate
possible peaks in the distribution, corresponding to matter overdensities,
which are far from the photo-$z$-determined position.

The upper panel shows that massive halos are very well localized by this
technique.  Such halos are likely to be tracing protocluster regions at
$z=2.5$.  The middle panel shows that lines of sight can cross multiple
overdensities (e.g.~crossing multiple filaments or protocluster regions
along the line of sight) and thus $p(z)$ can have more than one peak.
This is different from the very low $z$ case \citep{Ara15}, where there
are very few filaments or overdense regions within the survey to cause
confusion.
In the lower panel we show the difficulties in finding lower mass halos.
As Table \ref{tab:fractions} shows, the abundance of lower mass halos is
less sensitive to overdensities measured on $2.5\,h^{-1}$Mpc scales and
such halos do not produce a large (smoothed) flux decrement.
Thus only a very weak peak is seen at the true location of the halo, and
similar peaks are seen at many other locations.

In Fig.~\ref{fig:quantiles} we show a summary statistic for the error
rate of the predicted photo-$z$s for all of the lines of sight to
halos above $2\times 10^{12}\,h^{-1}M_\odot$ in the simulation at
$z=2.5$.  The plot shows the fraction of halos whose redshift is
correctly predicted, i.e.~the fraction of halos whose true redshift is
within the credible redshift region determined from the redshift PDF 
$p(z)$ computed with \eqq{pcondfit} 
(corresponding to grey regions in Fig.~\ref{fig:skewers}).  This is
shown as a function of the confidence level used to predict these
credible redshift regions.  As we can see, for halos above
$10^{13}\,h^{-1}M_\odot$, 90 per cent of the halos lie within the
50-per-cent-confidence-level credible region (i.e.~smaller than
$1\,\sigma$ for a Gaussian), while only 3 per cent of randomly placed
objects do.  The situation is less good for lower mass halos, but
still far better than random.  To further decrease the error rate, we
can choose a more conservative confidence level. For
example, if we choose to believe the 90-per-cent-confidence-level
credible regions, 99.5 per cent of the halos above
$10^{13}\,h^{-1}M_\odot$, 98.5 per cent of the halos above $5\times
10^{12}\,h^{-1}M_\odot$, and 92 per cent of the halos above $2\times
10^{12}\,h^{-1}M_\odot$ are predicted correctly.  By selecting
regions more likely to host massive halos the flux field is
drastically improving photo-$z$s.

Although a more conservative confidence level lowers the error rate of
predicted halo redshifts, it comes at the cost of increasing the
redshift uncertainty by broadening the credible redshift regions.
This can be seen in Fig.~\ref{fig:skewers}, where the conservative
90-per-cent-confidence-level credible regions (light grey) are broader
than those for the less conservative 68 per cent confidence level
(dark grey).  
For our fiducial flux smoothing scale of $R=2.5\,h^{-1}$Mpc
we find that the total width of the 90 per cent credible
region is $30-40\,h^{-1}$Mpc depending on halo mass. 
\changed{This is roughly the scale of large voids at the high redshifts we are probing.} 
\changed{This redshift uncertainty}
corresponds to $\Delta
z/(1+\bar z)\simeq 0.01$ at $\bar z=2.5$, which 
is significantly better
than typical photo-$z$ uncertainties (see Section~\ref{se:intro}).

Further improvements are possible if we had access to the flux
smoothed on smaller scales.  We demonstrate this in
Fig.~\ref{fig:credible_R}, which shows the redshift uncertainty given
by the total width of the credible region as a function of the flux smoothing
scale $R$.  With high-resolution flux fields smoothed on $R\sim
1\,h^{-1}$Mpc the redshift uncertainty could be reduced to
$10\,h^{-1}$Mpc, while the redshift error rate is essentially
independent of the flux smoothing scale (not shown).  Longer and more
expensive observations to obtain such high-resolution flux fields
could thus tighten redshift credible regions by a factor of a few,
without increasing the error rate of the redshift predictions
\citep[see][for a discussion of the observational requirements]{Lee14a}.
\changed{This would be important for mapping the cosmic web and 
for environmental studies.}

A potential caveat of our method is that the Ly$\alpha$ forest only
traces about $400\,$Mpc along the line of sight, so that a galaxy
could have higher (lower) redshift than the highest (lowest) redshift
for which we have any flux information.  Our method might then make a
false positive mistake by erroneously assigning a redshift inside the
region traced by the Ly$\alpha$ forest.
In our simulation, fewer than $10$ per cent of the galaxies that are
located at higher or lower redshift than the region traced by the
Ly$\alpha$ forest would be (erroneously) given a significant redshift
PDF inside the region traced by the Ly$\alpha$ forest.  Of course this
can be reduced further by trusting only the very highest peaks of the
redshift PDF.

\section{Future directions}
\label{sec:future}

There are numerous improvements that one could imagine to our simple
method.  First, we have treated all halos (above $10^{12}\,h^{-1}M_\odot$)
in the same manner.  If we knew in advance that a galaxy or its hosting
halo was particularly massive we could tune our selection to further improve
the performance by focusing on the most extremely overdense regions.
We tested this in simulations by calibrating the conditional probability to
find a halo given some flux using only halos above $10^{12.5}\,h^{-1}M_\odot$
or $10^{13}\,h^{-1}M_\odot$.  For the same halos, this tightens the 
total width of credible redshift intervals, but it also leads to a higher
fraction of objects outside the high-confidence interval.  Usually the
redshift error remains small, and it is outside the confidence interval because
that interval shrinks so much.
Whether this is an improvement depends on the ultimate application and
should be studied more quantitatively in the future.

Second, we have not included information about the shape of the cosmic web
in our analysis.  \citet{LeeWhi16} showed that Ly$\alpha$ forest tomography
is capable of classifying the observed volume into voids, sheets, filaments
and knots with high fidelity.  This could improve our recovery further,
\changed{e.g.~by assigning massive galaxies a higher probability to reside
in knots rather than filaments or sheets.}
As a simple first step in that direction, we \changed{characterize 
different structures of the cosmic web by smoothing 
the flux on two different smoothing scales,
noting that generically both smoothed flux fields should peak for clusters,
whereas only the high-resolution flux field might peak if a filament or sheet
crosses the sightline.
Based on this intuition we generalize our algorithm to }
use the conditional probability of finding a halo given the flux smoothed on
two different smoothing scales, $\delta_F$ and $\delta_F'$:
\begin{equation}
  P({\rm halo}|\delta_F,\delta_F') = \frac{P(\delta_F,\delta_F'|{\rm halo})}
{P(\delta_F,\delta_F')}.
\end{equation}
We compute this in simulations as a ratio of 2d histograms and fit this with
a multivariate Gaussian of the form
\begin{equation}
  P({\rm halo}|\delta_F,\delta_F') \propto
e^{-a\delta_F-a'\delta_F'-b\delta_F^2-b'\delta_F'^2
-c\delta_F\delta_F'}.
\end{equation}
We find that this slightly improves the total widths of credible redshift
intervals and the error rate, at the cost of making the
method somewhat more complicated.
It is possible that including the measured shear could further improve the
performance of the method, though properly characterizing the multivariate
probability distribution becomes more difficult.  
\changed{Another possibility for improvement might be to increase the integration
time of spectra, which would give higher resolution along the line of sight at 
the expense
of observing fewer sightlines at fixed total observation time, effectively reducing
the resolution perpendicular to the line of sight.}
We have not attempted to
implement these ideas here, as the simplest method is already performing
quite well.

Finally, we have treated each galaxy separately whereas one could imagine
a simultaneous recovery of the $p(z)$ for the entire population.  Such a
procedure is closer in spirit to the one in \citet{JasWan12} and may yield
dividends.  Such a forward modeling approach would allow us to take into
account the effects of redshift space distortions, bias and the cosmic web
using theoretical expectations of how galaxies and the IGM behave in a
model based on gravitational instability.
\changed{This method can also be combined with traditional photo-z's 
and alternative 
redshift estimation techniques like the ones presented in \citet{Ara15} and
in our paper.  }

Another interesting question is related to observational programs for
photo-$z$ calibration.  Typically, accurate spectra are measured for
all galaxies within the field of interest.  In contrast, with our
method one could imagine measuring only spectra for the brightest
background galaxies, and then using tomographic information from the
Ly$\alpha$ forest along the line of sight to calibrate photo-$z$s for multiple
galaxies along the line of sight -- independent of the galaxy spectral type
or whether it has prominent spectral lines.  From a single background
galaxy spectrum one could then calibrate photo-$z$s in the whole
$400$Mpc region along the line of sight. \changed{Based on the halo mass 
dependence of our results this should work best with very massive background
halos.
}

\section{Conclusions}
\label{sec:conclusions}

Determining the redshift of large numbers of cosmological objects is one
of the most difficult problems in observational cosmology.
In this paper we have shown that a smoothed map of the intergalactic medium,
obtained from spectra of distant galaxies, can be used to improve the
redshift accuracy of galaxies and AGN within hundreds of Mpc of the source.
This method works because in hierarchical structure formation the halos
which host galaxies and AGN preferentially live in the overdense regions
with small volume filling fraction.

Our method is extremely simple, once a map of the IGM has been obtained.
We use a simple Gaussian form for $p({\rm halo}|\delta_F)$ to transform
the observed flux perturbation, $\delta_F$, into a redshift PDF along the
line-of-sight to any galaxy.  This process can be repeated galaxy by galaxy.
Since the mapping between $\delta_F$ and $p(z)$ is monotonic, it is easy
to account for errors in the IGM map.

In the form presented herein, the method works best for the most massive
galaxies which live in the most massive halos, which tend to form in rare
regions of very negative $\delta_F$.  For such massive halos the redshift
accuracy and error rate are excellent: at our fiducial
$2.5\,h^{-1}$Mpc smoothing $\Delta z/(1+z)\simeq 0.01$ and 90 per cent of
halos above $10^{13}\,h^{-1}M_\odot$ lie within the 68 per cent credible
region.  Increasing the resolution of the IGM map reduces the redshift
uncertainty, while decreasing the resolution increases the uncertainty.
Lower mass halos demand a higher signal-to-noise, less smoothed map of
the IGM.  This is observationally more challenging.

Our method is extremely straightforward, but does not exhaust the information
available in IGM tomography.  By using more of the available information about
the cosmic web, and by performing a global reconstruction rather than by
analyzing galaxies one at a time, we expect to be able to work to lower masses
and further improve redshift performance.  We defer further developments of a
global analysis to future work.

\vspace{0.1in}

We thank Brice Menard and KG Lee for useful discussions.
The simulation, mock surveys, and reconstructions discussed in this work were
performed at the National Energy Research Scientific Computing Center,
a DOE Office of Science User Facility supported by the Office of Science of
the U.S. Department of Energy under Contract No.  DE-AC02-05CH11231.
This research has made use of NASA's Astrophysics Data System and of
the astro-ph preprint archive at arXiv.org.

\bibliographystyle{mnras}
\bibliography{ms}

\label{lastpage}
\end{document}